\documentstyle[12pt]{article}
\newcommand{\be}{\begin{equation}}
\newcommand{\ee}{\end{equation}}
\begin{document}
\pagestyle{empty}
\title{The phase diagram of the anisotropic  \\ Spin-1 Heisenberg Chain}

\author{A. L. Malvezzi \ \ and \ \ F. C. Alcaraz  \\
	 Departamento de F\'\i sica \\
	Universidade Federal de S\~ao Carlos \\
	 13565-905, S\~ao Carlos, SP, Brasil }
\date{KEY WORDS: Haldane gap, exact diagonalization, \\
conformal field theories, density matrix renormalizaton group}

\maketitle
\newpage
\pagestyle{plain}

In 1983, Haldane conjectured that the isotropic antiferromagnetic spin-S
Heisenberg chain has a gap or not depending if the spin is an integer or
half-integer\cite{haldane}. This conjecture is today well established
\cite{a,b}. In the case where an exchange z-anisotropy is present the
situation is less clear and the location of the point
where the massive Haldane phase
appears is still controversial even for the spin 1 case.
The Hamiltonian in
this last case is the spin-1 XXZ chain
\be
\label{H}
{\cal H} = \sum_{j=1}^{L} \ \left( S_{j}^{x}S_{j+1}^{x}
+ S_{j}^{y}S_{j+1}^{y} + \lambda S_{j}^{z}S_{j+1}^{z} \right)
\ee
where $S_{j}^{x},\ S_{j}^{y}$ and $S_{j}^{x}$ are
 the spin 1 operators.

{}From finite-size studies ( for lattices sizes up to $L=14$ )
and by exploring the
results of conformal invariance\cite{cardy}
Alcaraz and Moreo\cite{moreo}
conjectured that for $\lambda < \lambda_{c} = 0$ the Hamiltonian (\ref{H})
is in a massless X-Y phase with critical
 fluctuations governed by a conformal
field theory with central charge $c=1$. Moreover in the critical phase
$(-1\leq \lambda \leq 0)$ the dimensions of
operators $x_{n,m}^{(b)}, \ x_{n}^{(s)}$
governing the correlation functions in the bulk and in the surface are
given by\cite{moreo}
\be
\label{xnm}
x_{n,m}^{(b)} = n^{2}x_{p} + \frac{m^{2}}{4x_{p}} \, ,
\ \ \ \  \ x_{n}^{(s)} = 2x_{p}n^{2}
\ee
where
\be
\label{xp}
x_{p} = \frac{\pi - \cos^{-1}(\lambda)}{4\pi} \, ,  \ \ \ n,m = 0,\pm 1,
\pm 2, ... .
\ee

The Haldane phase is expected to appear when the
operator $x_{0,1}^{(b)} =
\frac{1}{4x_{p}} = \frac{\pi}{\pi - \cos^{-1}(\lambda)}$ becomes
relevant, which according to eqs. (\ref{xnm}) and (\ref{xp})
 occurs at $\lambda =
\lambda_{c} = 0$. This fact explains\cite{moreo} some exact degeneracies
appearing in the spectra of (\ref{H}) with free ends and is
corroborated by the results of ref. \cite{hatsu}.

More recently Yajima and Takahashi\cite{japa} by
calculating the spectra of
(\ref{H}) for lattice sizes up to $L=16$
 concluded that the Haldane phase
should starts at $\lambda = \lambda_{c} = 0.068\pm 0.003$. Since this
result destroy the conjecture (\ref{xnm})
we decide to make a more precise
calculation of $\lambda_{c}$ by using the density
matrix renormalization
group (DMRG) introduced by White\cite{white1}, which enable us to
 make spectral calculations in much
larger lattice sizes. The direct way to estimate $\lambda_{c}$ would be by
a direct mass gap evaluation of (\ref{H}) in a periodic chain. However our
results shows that for $\lambda \approx 0$ the finite-size gaps are very
small ( the transition at $\lambda_{c}$
 should be of Kosterlitz-Thouless type )
and it is very difficult to decide if the point
$\lambda = 0.065$ ( the worse
 estimative against the conjecture (\ref{xnm}) )
is in a massive phase or not. Another way to estimate $\lambda_{c}$ is
by calculating the exponents $\eta_{x}$, governing the correlation
$\langle S_{i}^{x}S_{i+r}^{x}(r) \rangle \sim r^{-\eta_{x}}$. The critical
phase disappears when $\eta_{x}$ reachs the value\cite{japa,chucrute}
$2x_{1,0}^{(b)} = 1/4$ or the surface exponent\cite{moreo}
$x_{1}^{(s)}$ reachs the value $1/4$.

Since the DMRG is much  more precise for spectral
calculations in open chains
we calculate $x_{1}^{(s)}$ in order to estimate $\lambda_{c}$.
This is done by
using the finite-size relations coming from the conformal
invariance of the
infinite system\cite{cardy}. The exponent $x_{(1)}^{s}$ is obtained  by
extrapolating $(L \rightarrow \infty)$ the sequence
\be
\label{x1}
x_{1}^{(s)}(L) = \frac{E_{1}^{1}(\lambda,L) - E_{1}^{0}(\lambda,L)}
{E_{2}^{0}(\lambda,L) - E_{1}^{0}(\lambda,L)}.
\ee
where $E_{j}^{n}(\lambda,L)$ is the $j^{\mbox{\underline{th}}}$ energy
in the sector where $\sum S_{i}^{z} = n$ of the Hamiltonian (\ref{H})
with free ends.

In table 1 we show for the special cases $\lambda = 0$ and
$\lambda = 0.065$ the estimated values for $x_{1}^{(s)}(L)$ for
lattices up to $L=48$. In our DMRG calculations we
 keep $k=$\ 50 and 80 states in
the truncated Hilbert space\cite{white1}. A good estimator for the errors
is $1-P_{k}$, where $P_{k}$ is the trace of the truncated density matrix.
The values in table 1 were obtained by taking $P_{k} \rightarrow 1$.

At $\lambda = \lambda_{c}$, the operator $x_{0,1}^{(b)}$, which corrects
the finite size scaling\cite{moreo}, should be marginal and logarithmic
corrections in the sequence (\ref{x1}) are expected. In fig. 1 we show the
data of table 1, at $\lambda = 0$, together with a mean square fit
in the form
\be
\label{x1fit}
x_{1}^{(s)}(L) = \frac{1}{4} + a \left( \ln L \right)^{b}.
\ee
showing a good numerical fit. A similar polynomial fit, which would be
the case if $\lambda_{c} > 0$, although possible shows larger deviations.
On the other hand the data of table 1 at $\lambda = 0.065$ are not well
 fitted by (\ref{x1fit}). As we can see from table 1 the data at $\lambda =
0.065$ clear indicate a limiting value $x_{1}^{(s)}(\infty)$ bigger than
1/4. In order to see this more clearly we calculated at $\lambda = 0.065$
 this estimator for $L=60$ sites and obtain $x_{1}^{(s)}(60) = 0.25023$,
which indicates that $\lambda = 0.065$ is
 already a point in the Haldane phase.

In conclusion, our results indicate that, in agreement with the conjecture
(\ref{xnm}),
the Haldane phase appears for $\lambda > \lambda_{c} = 0$,
and the logarithmic corrections appearing at this point explains the
numerical controversial in earlier numerical calculations.
\begin{center} {\bf Acknowledgments } \end{center}

This work was supported in part by CNPq and Funda\c c\~ao de Amparo \`a
Pesquisa do Estado de S\~ao Paulo - FAPESP, SP-BRASIL.

\newpage
\Large
Figure Caption
\normalsize
\vspace{1.0cm}

Fig. 1 - The data of table 1 together with the mean square fit
( continuum curve ).
\vspace{4cm}

\Large
Table Caption
\normalsize
\vspace{1.0cm}

Table 1 - Finite size estimatives of $x_{1}^{s}$, obtained from eqs.
(\ref{x1}) for $\lambda = 0$ and $\lambda = 0.065$.
\newpage
\Large
Table 1
\normalsize
\vspace{1.cm}

\begin{tabular}{||l|l|l||}	\hline
$L$	& $\lambda=0.0$ & $\lambda=0.065$ \\ \hline
8	& 0.236462774	& .246639580 \\
16	& 0.237501420	& .247661670 \\
24	& 0.237964476	& .248398065 \\
32	& 0.238271248	& .248954979 \\
40	& 0.238500913	& .249401545 \\
48	& 0.238685182	& .249774106 \\ \hline
\end{tabular}
\end{document}